\documentclass[a4paper]{article}  
\usepackage{multicol,lipsum}
\usepackage{graphicx}
\usepackage{a4wide}
\usepackage{txfonts}
\usepackage{amssymb}
\usepackage[T1]{fontenc}
\usepackage[utf8]{inputenc}
\usepackage{authblk}
\usepackage{natbib}
\usepackage{mathrsfs}
\usepackage{rotating}
\usepackage{setspace}
\usepackage{lscape}
\usepackage{color}

\providecommand{\keywords}[1]{\textbf{\textit{Keywords --}} #1}

\begin{document} 
\title{Ground-based detection of a cloud of methanol from Enceladus: When is a biomarker not a biomarker?}
\author[1,2]{\normalsize E. Drabek-Maunder\thanks{Emily.Drabek-Maunder@astro.cf.ac.uk}}
\author[1]{J. Greaves}
\author[3]{H. J. Fraser}
\author[2]{D. L. Clements}
\author[2]{L.-N. Alconcel}
\affil[1]{School of Physics and Astronomy, Cardiff University, Cardiff CF24 3AA, UK.}
\affil[2]{Imperial College London, Blackett Lab., Prince Consort Rd, London SW7 2AZ, UK.}
\affil[3]{School of Physical Sciences, The Open University, Walton Hall, Milton Keynes MK7 6AA, UK.}
\date{}
\maketitle   
\vspace{-8ex}


 \doublespacing
 \begin{abstract}
 
Saturn's moon Enceladus has vents emerging from a sub-surface ocean, offering unique probes into the liquid environment.  These vents drain into the larger neutral torus in orbit around Saturn.  We present a methanol (CH$_3$OH) detection observed with IRAM 30-m from 2008 along the line-of-sight through Saturn's E-ring.  Additionally, we also present supporting observations from the \emph{Herschel} public archive of water (ortho-H$_2$O; 1669.9 GHz) from 2012 at a similar elongation and line-of-sight.  The CH$_3$OH 5(1,1)-4(1,1) transition was detected at 5.9$\sigma$ confidence.  The line has 0.43 km/s width and is offset by +8.1~km~s$^{-1}$ in the moon's reference frame. Radiative transfer models allow for gas cloud dimensions from 1750~km up to the telescope beam diameter $\sim73000$~km.  Taking into account the CH$_3$OH lifetime against solar photodissociation and the redshifted line velocity, there are two possible explanations for the CH$_3$OH emission:  methanol is primarily a secondary product of chemical interactions within the neutral torus that (1) spreads outward throughout the E-ring or (2) originates from a compact, confined gas cloud lagging Enceladus by several km~s$^{-1}$.   We find either scenario to be consistent with significant redshifted H$_2$O emission (4$\sigma$) measured from the {\emph{Herschel}} public archive.  The measured CH$_3$OH:H$_2$O abundance ($>0.5$\%) significantly exceeds the observed abundance in the direct vicinity of the vents ($\sim0.01$\%), suggesting CH$_3$OH is likely chemically processed within the gas cloud with methane (CH$_4$) as its parent species.


\end{abstract}

%
\section{Introduction}

The discovery of liquid water below the icy surfaces of several moons orbiting Jupiter and Saturn is an exciting prospect for complex chemical and even biological activity taking place within them.  However, accessing these subsurface oceans is problematic. The subsequent discovery \citep{2006Sci...311.1406D, 2006Sci...311.1422H, 2006Sci...311.1401S, 2006Sci...311.1389K} of water vapour plumes venting from Saturn's moon Enceladus ($R_\mathrm{Enc} \sim 250$~km) through geyser-like structures near the moon's south pole \citep{2006Sci...311.1393P} , can help probe the interior processes of this particular object.

Most work to date on the molecular content of material ejected from Enceladus has been conducted through in situ observations by the {\em Cassini} spacecraft and by its Ion and Neutral Mass Spectrometer (INMS; \citealt{2006Sci...311.1419W,2009Natur.460..487W}).  Molecules like water, carbon dioxide, methane, methanol, ammonia and formaldehyde have been detected in the plumes; a further component of molecular mass 28 (CO or N$_2$) has also been seen. These observations are conducted in flybys at different altitudes above the surface and on different trajectories, so the properties of the entirety of the plume and any chemical processing within the plume (e.g. by Solar ultraviolet light) must be inferred.  

On larger scales, a neutral OH torus orbiting Saturn was found by \citet{1993Natur.363..329S}, which is fed by these active H$_2$O plumes  \citep{2001Icar..149..384J, 2005JGRA..110.9220J}.  The neutral torus is assumed to be centred on Enceladus' orbit (3.95~Saturn radii or $R_\mathrm{S}$; $R_\mathrm{S}=60268$~km), extending from 2.7 to 5.2~$R_\mathrm{S}$ \citep{2009Icar..202..280F}.  H$_2$O in the neutral torus has been further explored by the \emph{Herschel} Space Observatory \citep{2011A&A...532L...2H}, while low signal-to-noise has limited \emph{Cassini} from performing in situ measurements.

The \emph{Cassini} mission is nearing its end and conducted its final flyby of Enceladus at the end of 2015. With little or no prospect of a new mission to Saturn before 2030, further studies and monitoring of the Enceladus plume must be done remotely from Earth.  Fortunately, submillimetre spectroscopy is well suited to such studies since many of the organic molecular species of interest have transitions at these wavelengths.  A disadvantage is that single-dish telescopes, while suited to temporal monitoring, trace larger regions than the size of the plumes.  In light of further developments in the plume composition from \emph{Cassini} and neutral gas environment surrounding Enceladus by \emph{Herschel}, we present the first results from a programme of submillimetre spectroscopic observations of a gas cloud near Enceladus using ground-based observatories from early 2008.

This paper is organised as follows: Section~\ref{section_observations} details the ground-based observations targeting methanol (CH$_3$OH) and Section~\ref{results} presents the methanol spectrum.  Section~\ref{models} details the radiative transfer and dynamical models used to constrain the methanol abundance and the likely region from which the methanol originates.  Lastly, we summarise our results and discuss their implications for using methanol as a biomarker in solar system objects and exoplanet environments in Section~\ref{conclusions}. 

\section{Observations}
\label{section_observations}

\begin{figure}
\centering
   \includegraphics[width=3in]{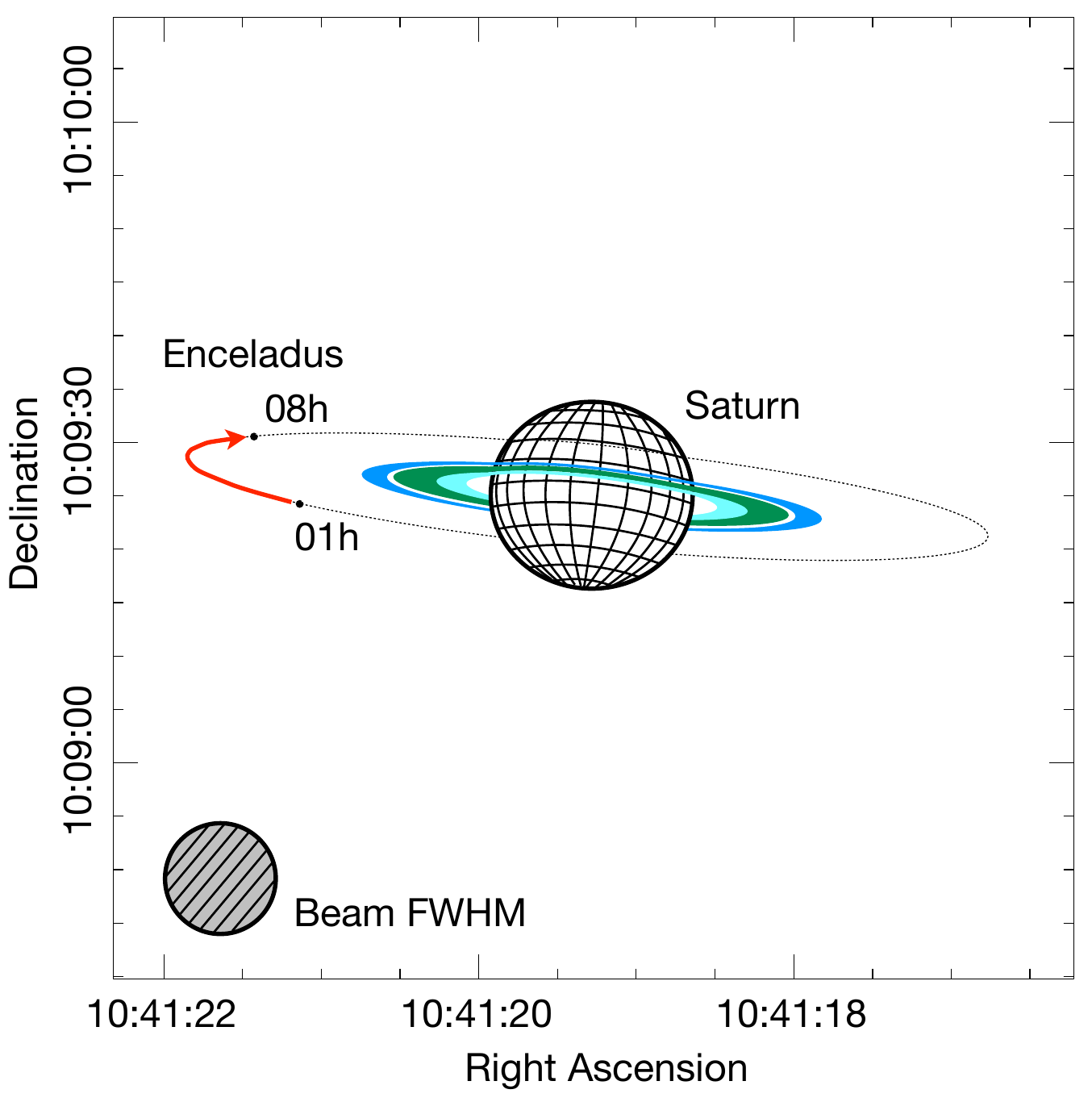}
   \caption{Location of Enceladus and Saturn during the CH$_3$OH observations on 10 Jan 2008 (01-08~hours UT).  The C- (innermost ring; light blue), B- (centre ring; green) and A-rings (outer ring; blue) are shown, along with the centre of the E-ring (dotted line), which also corresponds to Enceladus' orbital radius.}
   \label{Enc_location}%
    \end{figure}

The observations were made at the Instituto de Radioastronomie Milimetrica (IRAM) 30~m telescope 
at Pico Veleta, Spain at 01-08~hours UT on 10 Jan 2008 using the HERA 
camera and the VESPA spectrometer. The CH$_3$OH~5(1,1)-4(1,1) line at 239.7463~GHz 
was observed 
with 80~MHz passbands with a telescope beam size of $\sim$10.5$''$ FWHM. 
The observations were made by frequency-switching over a narrow 3.45 MHz interval to maintain flat baselines using 80 kHz channels in VESPA. 
Figure~\ref{Enc_location} shows the location of Enceladus w.r.t. Saturn at the time of observations.  
The angular size of Enceladus was 0.08$''$ with Saturn 8.6~AU from the 
Earth. The ring opening angle was -6.9$^\circ$.  We note the only possible contaminant within the beam was the moon Mimas towards the end of the observing period.

The telescope could track Enceladus but the acquisition software was only able to track the velocity of Saturn.  Saturn was offset by $\geq$~30$''$ during the observations to prevent spectral contamination of the Enceladus data. Velocity shifts to place the data in the Enceladus rest-frame were then made in reduction software for each observation.  In total 48 observations were made, with the drift between observations being $<$0.1 km/s on average (worst case of 0.4 km/s). The applied shift varied from -9.2 km/s at the start to -12.6 km/s at maximum elongation followed by reversal back to -10.6 km/s at the end of the track. Hence, any artefacts associated with particular spectrometer channels would be smeared over a few km~s$^{-1}$ in the co-added data, and so not be able to create a false-positive narrow line.

Details of the data reduction (i.e. despiking, baseline-fitting and baseline-subtraction) are given in the Appendix (see Figure~\ref{fig:nobaseline}).  The spectra are automatically calibrated to a T$_\mathrm{A}^\ast$ antenna temperature scale.  However, we must correct the antenna temperature scale for telescope inefficiency.  The efficiencies were not measured during our run, but the HERA User Manual (V2.0; Nov 2009) stated the beam efficiency is 0.52 at 230~GHz and the forward efficiency is 0.90. We calculate a main-beam temperature T$_\mathrm{MB}$ = F$_\mathrm{eff}$/ B$_\mathrm{eff}$ T$_\mathrm{A}^\ast$, i.e. T$_\mathrm{A}^\ast$ is divided by 0.58 to give T$_\mathrm{MB}$.  

In addition to the CH$_3$OH spectrum, we also include publicly available observations taken by \emph{Herschel} (originally from project ID OT2\_elellouc\_3, PI Lellouch) of the ortho-H$_2$O $\mathrm{2_{12}}$--$\mathrm{1_{01}}$ line at 1669.9 GHz, taken on 27 June 2012 at $\sim22$:00~UT over a period of $\sim0.33$~h using the HIFI instrument.  The observations were taken in the fast dual beam switch (DBS) raster mapping mode, where we extracted the spectrum using only the observation centred directly on Enceladus.  The telescope beam FWHM is $\sim13''$ at $\sim1660$~GHz.  During this observation, Enceladus was at similar elongation from Saturn as the observations from IRAM, where the rings were edge-on.

\section{Results}
\label{results}

The co-added CH$_3$OH spectrum is shown in Figure~2a with 0.1~km~s$^{-1}$ velocity channels and in a frame co-moving with Enceladus.  The peak antenna temperature T$_\mathrm{A}^\ast$ is 0.038$\pm0.011$~K (3.6$\sigma$) and integrated intensity $\int \mathrm{T_A^\ast} \ \mathrm{dv}$ is 0.018$\pm0.003$~K~km/s (5.9$\sigma$).  This is the only significant feature within the passband, where all other signals summed across 0.8 km/s intervals (totalling 125 intervals across the 80~MHz passband) are $\leq3\sigma$ (i.e. $\leq0.009$~K~km~s$^{-1}$).  See the Appendix for more detail regarding the robustness of the CH$_3$OH detection (i.e. Figure~\ref{fig:nobaseline}).   Accounting for the beam and forwards efficiency, the peak main-beam temperature $\mathrm{T_{MB}}$ is 0.066~K and integrated intensity $\int \mathrm{T_{MB}} \ \mathrm{dv}$ is 0.031~K~km/s.

The line FWHM is 0.43 km/s as measured by fitting a Gaussian to the methanol detection (Figure~2), and the line is centred at +8.1~km/s 
relative to the moon. This indicates that we are not looking at material in the direct vicinity of the plume origin, but material that has been redshifted by some means.  

The ortho-H$_2$O $\mathrm{2_{12}}$--$\mathrm{1_{01}}$ line is shown in Figure~2b with 0.3~km~s$^{-1}$ velocity channels in a co-moving Enceladus frame.  The full line profile must be fitted with a two-component Gaussian due to a significant redshifted H$_2$O line-wing.  This redshifted emission spans $\sim$14 velocity channels (i.e. up to $\sim 7$~km~s$^{-1}$) with a total signal $\sim0.421 \pm 0.105$~K~km~s$^{-1}$ (i.e. 4$\sigma$).  Even though the venting process is variable \citep{2013Natur.500..182H}, the \emph{Herschel} H$_2$O observation is redshifted by a similar amount as the CH$_3$OH detection four years earlier.

   \begin{figure*}
   \centering
   \includegraphics[width=5in]{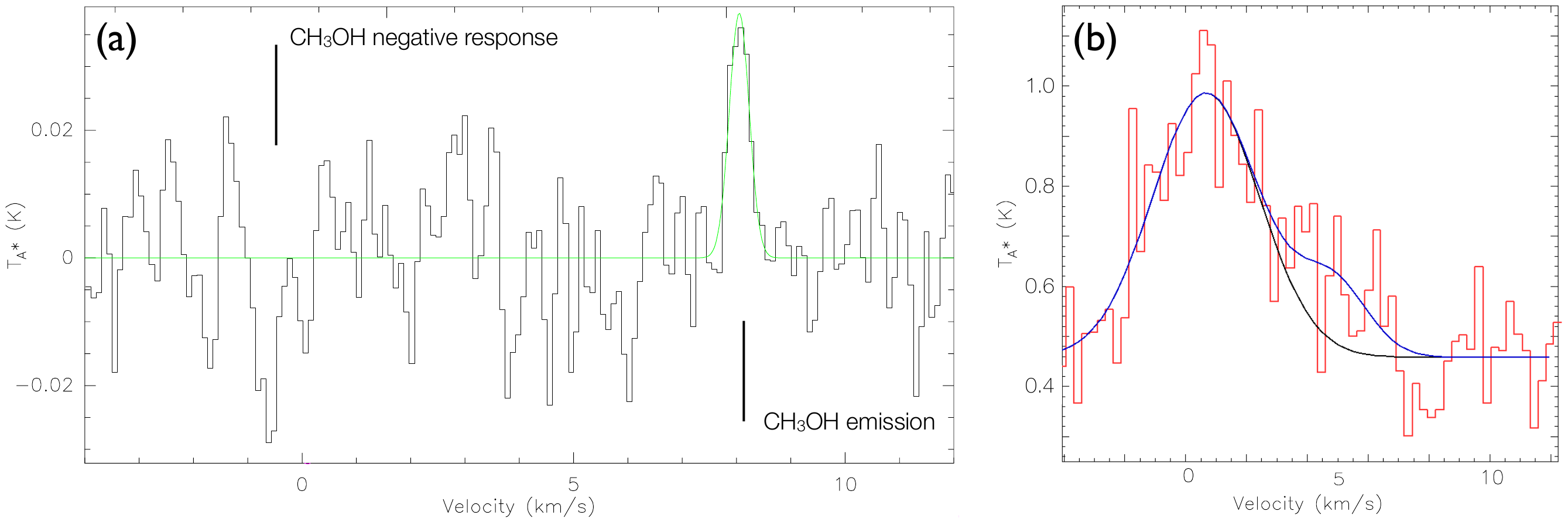}
   \caption{CH$_3$OH and H$_2$O observations of Enceladus in the moon's velocity frame.  \emph{Left:}  CH$_3$OH spectrum in antenna temperature T$_\mathrm{A}^\ast$ (K), where the Gaussian fit to the line is overplotted in green.  We note that the feature at $\sim -0.5$~km~s$^{-1}$ is the CH$_3$OH line seen as a negative response. The 3.45~MHz frequency-switching interval causes the CH$_3$OH line emission to appear as both a positive and negative response, separated by twice the interval size (i.e. 6.9~MHz or $\sim 8.6$~km~s$^{-1}$).  \emph{Right:} H$_2$O spectrum in antenna temperature T$_\mathrm{A}^\ast$ (K).  We fit a Gaussian to the main peak in the spectrum (black) and a combination of two Gaussians to fit the full line profile.}
   \label{FigSpectrum}%
    \end{figure*}

\section{Models}
\label{models}

Even though methanol has been found within the plume components \citep{2009Natur.460..487W}, the region in which IRAM can detect methanol must be on a larger scale.  The diameter of the telescope beam subtends $\sim$73,000~km ($\sim1.2 \ R_\mathrm{S}$), which is much larger than Enceladus' diameter of only $\sim$500~km.  Even if the methanol cloud is optically thick and warm (e.g. plume temperatures are $\sim180$~K; \citealt{2006ApJ...644L.137J, 2010Icar..209..696C}), the minimum size we are sensitive to is $\sim1750$~km, or 7~$R_\mathrm{Enc}$.  On the other hand, diffuse gas in the larger neutral torus could fill the beam (where the torus has a radial extent of $R_\mathrm{Enc}\pm1.25 \ R_\mathrm{S}$ and a vertical extent of $\pm0.4 \ R_\mathrm{S}$; \citealt{2009Icar..202..280F, 2011A&A...532L...2H} ).

For generality, we consider a cloud of methanol with length $l$, and use the non-local thermodynamic equilibrium (non-LTE) radiative transfer code \textsc{RADEX} to find solutions for the radiation temperature ($\mathrm{T_R}$).  Details of \textsc{RADEX} and the full parameter space of the models that fit the CH$_3$OH observation can be found in Section~\ref{radex_explained}.  For clouds with size scales smaller than the telescope beam (i.e. $l<73,000$~km), we must account for beam dilution which causes the main-beam temperature to be lower than the expected radiation temperature by $\mathrm{T_{MB}= \eta_{dil} \ T_R}$, where $\mathrm{\eta_{dil}}= \frac{l^2}{l^2 + l^2_\mathrm{beam}}$ and $l_\mathrm{beam}$ is the beam diameter.

For the radiative transfer models, we set the gas kinetic temperature $\mathrm{T_{gas}} = 100$~K (e.g. \citealt{2011A&A...532L...2H}) and the line FWHM at 0.43~km~s$^{-1}$ as observed (Section~\ref{results}).  While plume temperatures reach $\sim180$~K \citep{2006ApJ...644L.137J, 2010Icar..209..696C}, we expect plumes to cool to Saturn's thermal field ($100$~K) at the scales traced by our observations (see \citealt{2011A&A...532L...2H} for modelling of the gas temperature within the neutral torus).  From the gas kinetic temperature, we can calculate the non-thermal (turbulent) velocity dispersion of methanol by $\mathrm{\sigma_{NT}^2 = \sigma_{CH_3OH}^2 - \sigma_T ^2}$, where $\mathrm{\sigma_{NT}}$ is the non-thermal line width, $\sigma_\mathrm{CH_3OH}$ is the observed line width of CH$_3$OH (calculated from $\mathrm{\sigma_{CH_3OH}= v_{FWHM} / \sqrt{8\ln(2)}}$) and $ \mathrm{\sigma_T}$ is the thermal component of the line width (assuming a 100~K cloud).  Removing the thermal component leaves a turbulent velocity dispersion of $\sim0.1$~km~s$^{-1}$, suggesting the methanol gas is dispersing at $\leq0.1$~km~s$^{-1}$.  

Solutions for $\mathrm{T_{MB}}$ were sought for different CH$_3$OH:H$_2$O abundance ratios ($X$), where we relate the H$_2$O density, $\mathrm{n(H_2O)}$, to the CH$_3$OH column density, $\mathrm{N(CH_3OH)}$ by $\mathrm{n(H_2O) = \frac{N(CH_3OH)}{\emph{l} X}}$.  The CH$_3$OH:H$_2$O abundance is constrained as $X\leq 5$\%, which is the expected C:H$_2$O budget (estimated from the CO$_2$:H$_2$O abundance within Enceladus' plumes; \citealt{2009Natur.460..487W}).  If methane CH$_4$ is the parent species of CH$_3$OH, then the preferred CH$_3$OH:H$_2$O abundance is $X\leq1$\%.  In \textsc{RADEX}, the collision partner is H$_2$, so we set density to be 10$\times$ higher to account for the higher mass of the actual collision partner H$_2$O.

Additionally, we find solutions for the cloud length $l$ that reproduce $\mathrm{T_{MB}(CH_3OH)}$ within the allowed CH$_3$OH:H$_2$O abundance.  Furthermore, a maximum H$_2$O column density was applied.  The maximum assumes that all of the H$_2$O molecules vented (estimated to be $10^{28}$ molecules~s$^{-1}$; \citealt{2010JGRA..11510252S}) will survive during the photodissociation timescale $\sim2.5$~months (e.g. \citealt{2010Icar..209..696C, 2010JGRA..11510252S, 2011A&A...532L...2H}) within the specific region of size $l$.  

\begin{figure*}
\centering
   \includegraphics[width=5in]{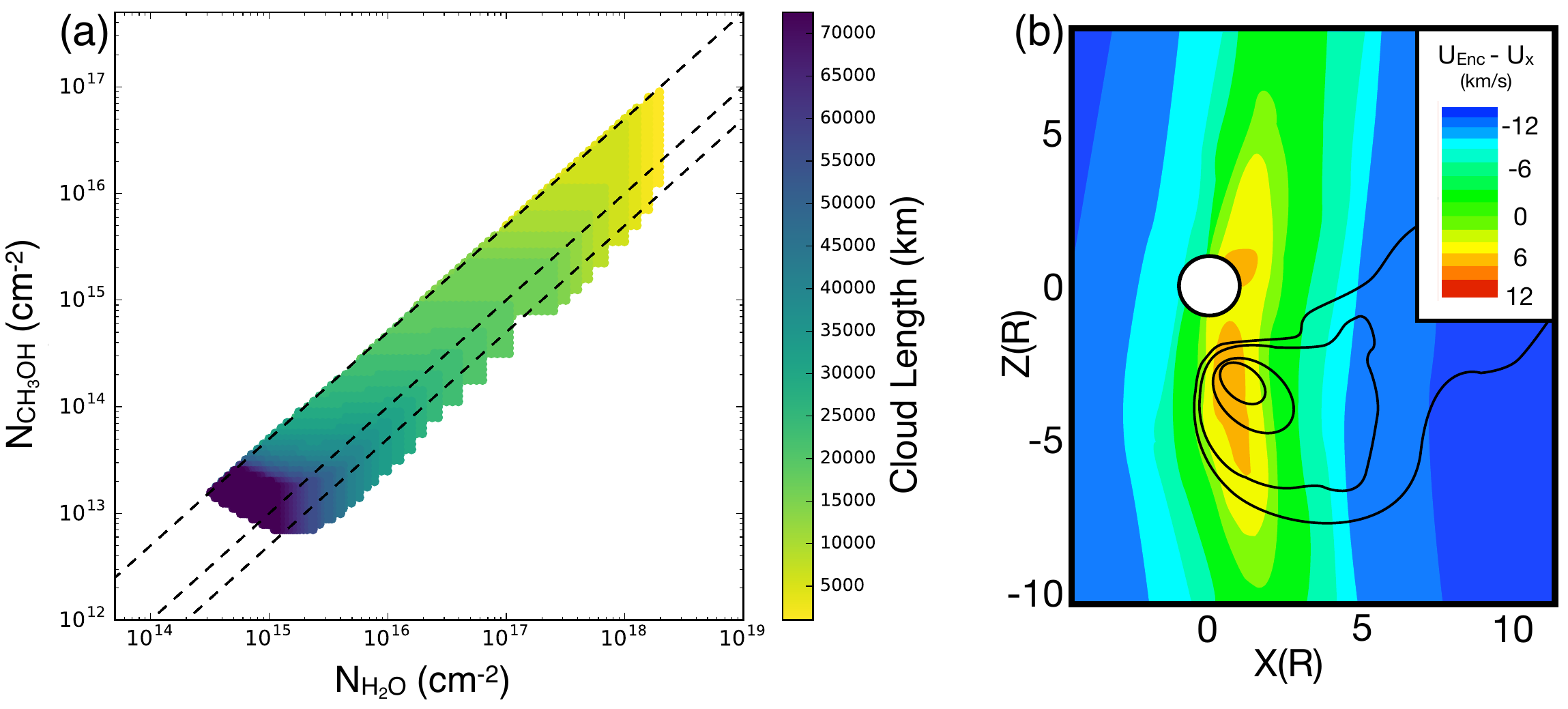}
   \caption{{\emph{Left:}}  \textsc{RADEX} model results for CH$_3$OH and H$_2$O column densities and varying cloud lengths that match the CH$_3$OH observation.  Colours correspond to cloud lengths 1750--75000~km (in increments of 250, 500 and 2500~km at ranges $l<2500$~km, $2500<l<5000$~km and $l>5000$~km respectively).  Dashed lines (left to right) correspond to CH$_3$OH-to-H$_2$O abundances of 5\%, 1\% and 0.5\%. {\emph{Right:}}  Plasma velocities near Enceladus, based on Figure~2 in \citet{2010JGRA..115.4215J}.  The component X (R$=R_\mathrm{Enc}$) is along the direction of the corotational flow around Saturn and component Y is positive towards Saturn.  Velocities are measured w.r.t. Enceladus' orbital velocity.  Contours correspond to the highest plasma densities, $\mathrm{n}\sim1$--4$\times10^{2}$~cm$^{-3}$.}  
   \label{models_radex}%
    \end{figure*}

Figure~\ref{models_radex} depicts the models that match the expected CH$_3$OH radiation temperature (i.e. the main-beam temperature corrected for beam dilution) within a conservative 50\% uncertainty for each cloud length $l$.   We investigate the CH$_3$OH:H$_2$O abundance by comparing the methanol and water column densities, $\mathrm{N(CH_3OH)}$ and $\mathrm{N(H_2O)}$ for each cloud length.  We find allowed CH$_3$OH cloud dimensions $l$ to be 1750--72500~km.  At smaller cloud lengths ($l<1750$~km), methanol becomes increasingly beam diluted and high densities and column densities are needed to produce the methanol detection.  However, CH$_3$OH also becomes optically thick and higher gas temperatures are needed to adequately reproduce the beam-diluted detection.  We find the modelled CH$_3$OH-to-H$_2$O abundance is relatively independent of cloud length, ranging from 0.5--5\%.  This abundance is $\sim$50$\times$ higher than what is found in the direct vicinity of the plumes \citep{2009Natur.460..487W}, implying chemical processing has occurred.  

The cloud size scenarios are summarised in the Appendix (Table~\ref{full_table_parameters}), including the range of H$_2$O densities from the \textsc{RADEX} models that fit the CH$_3$OH observation. For smaller cloud lengths (e.g. $l=1750$~km), we have H$_2$O densities ranging from $\mathrm{n_{RADEX}(H_2O)}=1.5\times10^9$--1.1$\times10^{10}$~cm$^{-3}$.  This density range is reasonably in agreement with past \emph{Cassini} measurements of the E3 and E5 flybys, which found the peak H$_2$O density to reach $\sim10^8$--10$^9$~cm$^{-3}$ near the plumes (Teolis et al. 2010; Smith et al. 2010).  At larger cloud lengths (e.g. $l=72500$~km), we find the H$_2$O density range to be $\mathrm{n_{RADEX}(H_2O)} = 4.5\times10^4$--1.6$\times10^5$~cm$^{-3}$, slightly above \citet{2010Icar..209..696C}, which found neutral densities in the torus $\sim10^4$~cm$^{-3}$.  

We can estimate how far the CH$_3$OH is able to travel outward from Enceladus from the photodissociation timescales under solar irradiation.  If the gas is dispersing at $\leq0.1$~km~s$^{-1}$ and the photodissociation timescale is $\mathrm{t_{photo} = 1.7\times10^6}$~s (accounting for Saturn's distance at 9.3~AU; \citealt{1992Ap&SS.195....1H}), then we expect a methanol cloud extent up to $\leq173,000$~km (i.e. $\sim 3$~$R_\mathrm{S}$).  This means the methanol is easily able to spread into the neutral torus and fill the beam of the telescope.  We note that H$_2$O and CH$_4$ have even longer photodissociation timescales than CH$_3$OH and are not a limiting factor.


Therefore, there are two possible explanations for the narrow CH$_3$OH line observation:  methanol is made both by Enceladus and through other chemical pathways within the neutral torus that (1) spreads outward or (2) remains in a more compact, confined gas cloud that trails Enceladus' orbit by several km~s$^{-1}$.

The velocities at which methanol is seen are consistent with a torus model, in particular with molecules spreading outwards from Enceladus' orbit \citep{2009Icar..202..280F}.  At the time of observations, Saturn's rings were tilted such that a beam pointed at Enceladus' position passed preferentially through the far-side of the neutral torus and even broader E-ring (3-9~$R_\mathrm{S}$).  Molecules further out than Enceladus will orbit with slower speeds.  If we are observing the far-side of the torus or E-ring, molecular line observations will be redshifted w.r.t. Enceladus as it approaches.  This is seen in the observations in Figure~\ref{FigSpectrum}.  The magnitude of the line-of-sight velocity shift w.r.t. Enceladus can be fitted assuming the molecules are in Keplerian rotation in the ring plane, which was tilted w.r.t. Earth at 7$^\circ$ in 2008.  The detected methanol must be at an orbit of $\sim8$~$R_\mathrm{S}$ to be in the telescope beam, assuming most of the spectral emission was contributed around the time of maximum elongation.  However, methanol ejected directly from Enceladus through plumes may only be able to spread by $\sim 3 \ R_\mathrm{S}$ accounting for a methanol photodissociation time at the distance of Saturn (i.e. out to a radius $\sim7$~$R_\mathrm{S}$ accounting for the distance between Saturn and Enceladus; \citealt{1992Ap&SS.195....1H}).  Therefore, the detected methanol would have to be a secondary product of chemical processing within the larger neutral torus in this scenario, as discussed below.

The second possibility is that a compact, confined gas cloud trails Enceladus' orbit by several km~s$^{-1}$.  An example of a trailing gas cloud can be seen in a model by \citet{2010JGRA..115.4215J}, which was reconstructed from \emph{Cassini} flyby measurements of the plasma.  In Figure~\ref{models_radex}, we show a diagram of Enceladus with the plasma velocities and density contours (based on Figure~2 in \citealt{2010JGRA..115.4215J}).  The plume emerges from the southern pole of Enceladus (to -Z) and streams along the X-axis, which is in the line-of-sight for our observations.  Electrons south of Enceladus' plumes form a denser cloud at length $l\lesssim1750-2000$~km (7--8~R$_{\mathrm{Enc}}$) and trail Enceladus by $\lesssim6$~km~s$^{-1}$.  We have observed CH$_3$OH redshifted by a similar $+8.1$~km~s$^{-1}$ and over a similar-sized region.  However, further modelling and observations are needed to explain the mechanism in which gas would be confined by Saturn's magnetosphere.


Additionally, there is independent evidence of a significant (4$\sigma$; see Section~\ref{results}) redshifted molecular component from archival ortho-H$_2$O spectrum taken when Enceladus was at a similar elongation from Saturn.  Since the active plumes are thought to feed the neutral torus in orbit around Saturn, it is expected that H$_2$O will fill the \emph{Herschel} telescope beam and be present at both small and large scales surrounding Enceladus.  In particular, this redshifted H$_2$O component can result from gas spreading \citep{2009Icar..202..280F} or the presence of a confined cloud of gas lagging the orbital velocity of both the moon and the co-rotating neutral torus, where the latter process depends on the exact velocity of outgassing material at the time.  If either gas spreading or a confined cloud are the causes of the redshifted molecular emission, we do not necessarily expect the H$_2$O and CH$_3$OH observations to be redshifted by the same amount because H$_2$O may be tracing different radii than the CH$_3$OH detection and/or temporal variations observed from the plumes \citep{2013Natur.500..182H}.  Careful monitoring of Enceladus needs to be done in the future to better understand how the plume processes change over time.

Considering that water, hydrogen, oxygen and hydroxyl are abundant in the torus, the rate of methanol production may depend on the availability of methane, which is a $\sim1.6\%$ CH$_4$:H$_2$O abundance \citep{2009Natur.460..487W}, or other hydrocarbons.  Using public astronomical networks (KIDA\footnote{http://kida.obs.u-bordeaux1.fr}, UMIST; \citealt{2013A&A...550A..36M}), we suggest possible routes to methanol through gas-phase associate detachment: \begin{equation} \mathrm{CH_3 + OH^- \rightarrow CH_3OH + e^-} \end{equation}  While other routes for the creation of methanol are possible, these involve more complex molecules that are less abundant and may not be present in the plume environments.  

\citet{2010Icar..206..618C} find that dissociative electron attachment can produce negative water-group ions (e.g. OH$^-$, O$^-$, H$^-$) in Enceladus' plumes from neutral H$_2$O, where O$^-$ and H$^-$ are produced by polar photodissociation of H$_2$O by photons in $\sim36-100$~eV.  While H$^-$ has a relatively short photodissociation timescale ($\sim6$~seconds at Saturn; see \citealt{1992Ap&SS.195....1H}), \citet{2014M&PS...49...21C} find that H$^-$ quickly reactions with H$_2$O to produce $\mathrm{OH^- + H_2}$ in the comet 1P/Halley environment, where both OH$^-$ and O$^-$ are abundant within the coma ($\leq1000$~km from the nucleus).  This cometary environment may be similar to Saturn, where Saturn is a soft X-ray source (scattering X-rays from the Sun; \citealt{2007P&SS...55.1135B}) that could form these water-group ions.  

While the methyl radical is not directly modelled from the Cassini INMS data \citep{2006Sci...311.1419W, 2009Natur.460..487W}, it is a key molecule in the production of more complex hydrocarbons (e.g. Titan; \citealt{2006P&SS...54.1177A}) and is likely present within Enceladus' plumes or created once material is ejected \citep{2010Icar..206..618C}.  One possible pathway is through associative attachment from ethylene (CH$_2$) and water-group ions, where traces of ethylene have noted in the plume environment \citep{2010epsc.conf..305W}: \begin{equation} \mathrm{CH_2 + H^- \rightarrow CH_3 + e^-} \end{equation}  Similarly, reactions directly with methane and other molecules abundant in the plume environment can produce the methyl radical.  For example, ion-neutral reactions: \begin{equation} \mathrm{CH_4 + O^- \rightarrow CH_3 + OH^-} \end{equation} \begin{equation} \mathrm{CH_4 + H_2O^+ \rightarrow CH_3 + H_3O^+,} \end{equation} where H$_2$O$^+$ and H$_3$O$^+$ have been found directly (\citealt{2009GeoRL..3613203T}; altitudes $\sim200$~km).  Neutral-neutral reactions with methane are also possible with neutral water-group atoms and molecules present in the E-ring environment: \begin{equation} \mathrm{CH_4 + OH \rightarrow H_2O + CH_3} \end{equation} \begin{equation} \mathrm{CH_4 + O \rightarrow OH + CH_3} \end{equation}  

In all of these proposed scenarios, the plasma density appears to be an integral part of the gas-phase chemical production of CH$_3$OH and enhancement in CH$_3$OH abundance once the plume material is ejected from Enceladus.  This further supports the offsets in velocity we see in the CH$_3$OH spectrum and the plasma within 2000~km from the south pole of Enceladus \citep{2010JGRA..115.4215J}, where this material is travelling more slowly than the moon.  Plasma densities are found to peak at $\sim10^2$~cm$^{-3}$, indicating an electron abundance $\geq10^{-7}$ w.r.t. modelled H$_2$O (see Table~\ref{full_table_parameters}), which is similar to electron abundances found in the ISM (w.r.t. H$_2$; \citealt{2007ARA&A..45..339B}).  As stated above, further observations are needed to constrain possible variability in the plume and surrounding environment which may lead to changes in density, chemical composition and velocity of the ejected material over time.

We note that \citet{2009GeoRL..3617103H} finds a potential CH$_3$OH ice surface feature, though this can also be interpreted as hydrogen peroxide H$_2$O$_2$ \citep{2007ApJ...670L.143N}.  However, past work has also shown that CH$_3$OH tends to be destroyed in ice pathways, making it less likely that CH$_3$OH is produced from ice on Enceladus' surface \citep{2014A&A...570A.120B}.


%

\section{Discussion and Conclusions}
\label{conclusions}
We report on the first ground-based detection of a molecule ($\mathrm{CH_3OH}$) in Enceladus' plumes.  Radiative transfer models suggest the origin of methanol can be from a cloud with length ranging from 1750~km up to a size comparable to the neutral torus in orbit around Saturn.  There are two possible explanations for producing the redshifted CH$_3$OH and H$_2$O, taking into account the solar photodissociation timescale of CH$_3$OH and the observed redshifted CH$_3$OH emission.  First, methanol may be a secondary product of chemical processing within the neutral torus and spread further out into the E-ring to a radius $\sim 8 \ R_\mathrm{S}$ (e.g. \citealt{2009Icar..202..280F}) at velocities smaller than Enceladus' rest-frame velocity (i.e. due to Keplerian rotation).  Alternatively, methanol may original from a smaller, beam-diluted region nearer to Enceladus (i.e. $l<2000$~km).  To further distinguish between these two mechanisms, further modelling and observations are needed.



What is the origin of the methanol cloud surrounding Enceladus?  Since methanol is expected to be present in an organic environment \citep{2007Icar..187..569M}, an exciting prospect is that the observed methanol is being produced by living organisms within Enceladus' subsurface ocean.  In Earth's oceans, \citet{minceraicher} finds that methanol is produced at a rate as high as 0.3\% w.r.t. the total cellular carbon.  If we assume that methanol production within subsurface oceans on Enceladus is similar to Earth and all of the carbon measured in Enceladus' vents is expelled by microbes in the subsurface ocean (i.e. C:H$_2$O$\sim5$\%; \citealt{2009Natur.460..487W}), then we would expect a CH$_3$OH:H$_2$O abundance within Enceladus' subsurface oceans to be $\sim0.015$\% in an organic environment.  This result is similar to the CH$_3$OH abundance measured in the direct vicinity of the vents by \emph{Cassini} at $\sim0.01$\% \citep{2009Natur.460..487W}, indicating that it is possible for the CH$_3$OH found in the vents to be a biomarker in an extreme case with the carbon predominantly of biological origin.

In contrast at larger scales, our CH$_3$OH observation suggests the gas cloud trailing Enceladus has a CH$_3$OH:H$_2$O abundance that is $\sim$50$\times$ higher than in the vents.  Therefore, it is likely methanol is being produced once the material is ejected from the subsurface ocean, making it improbable as a  biomarker signature in this particular case.  In the future, caution should be taken when reporting on the presence of supposed biomarkers, in both solar system and exoplanet environments.  The most robust method for investigating the complex chemistry, particularly in subsurface ocean environments, is obtaining observations close to the vents.

\section{Acknowledgements}
ED acknowledges funding from Cardiff University.  This work uses observations from project number 220-07 with the IRAM~30m Telescope. IRAM is supported by INSU/CNRS (France), MPG (Germany) and IGN (Spain).  We would like to thank the referee Chris McKay for carefully reading this work.  Lastly, members of our team acknowledge the three birthdays lost to this project, for the pursuit of science.


\appendix

\section{Methods and detection overview}

\begin{sidewaysfigure}
   \centering
   \includegraphics[width=7in]{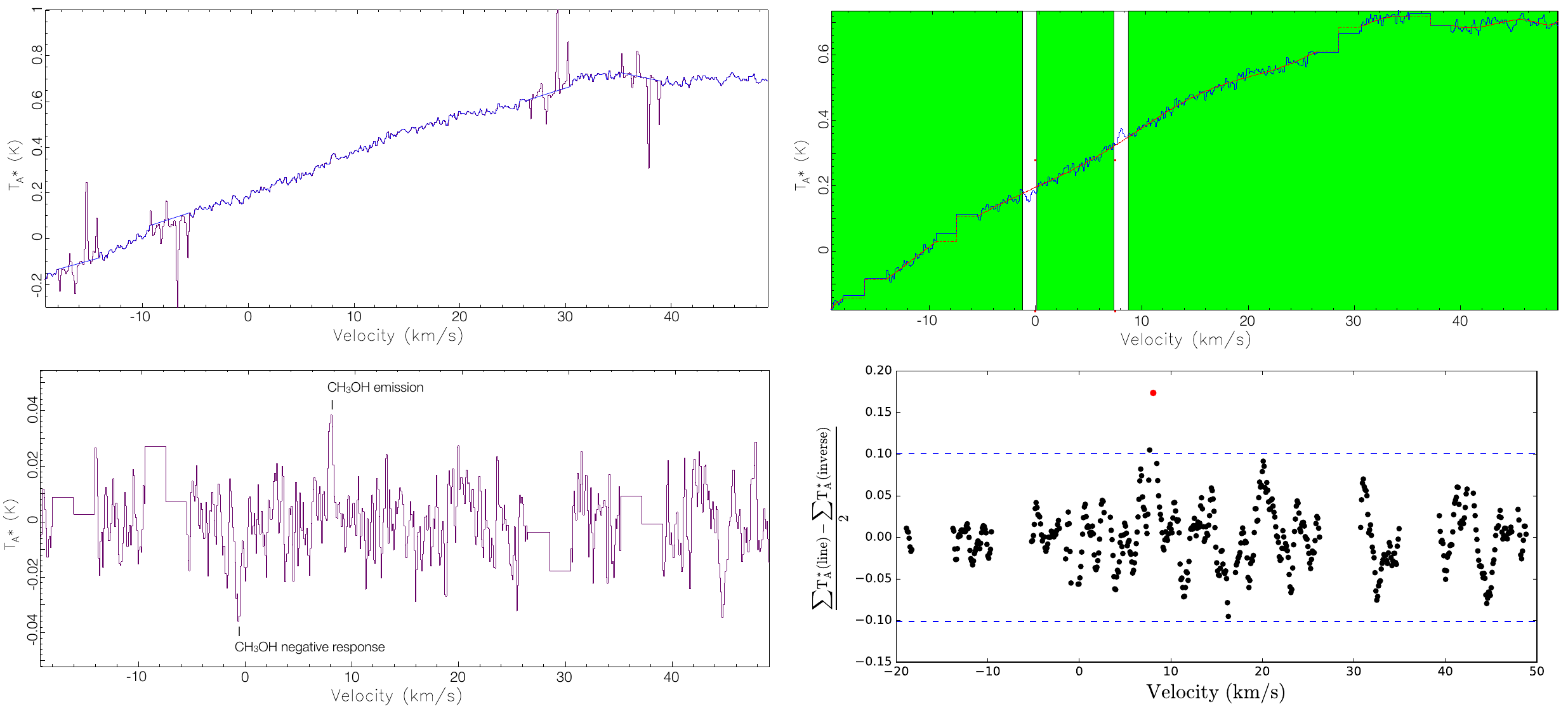}
   \caption{{\emph{Top Left:}}  Full spectrum from the IRAM 30-m telescope centred on 239.7463~GHz prior to fitting and subtracting the baseline.  The spectrum was produced by shifting individual observations in velocity to Enceladus' rest frame and then averaged. Noticeable spikes (in both positive and negative due to frequency-switching) can be seen at velocity channels towards the ends of the band.  {\emph{Top Right:}}  Baseline-fitting was done with a 14th order polynomial (the highest available in SPLAT) over the regions within the green boxes; the stepped parts show where channels are blanked in the spike removal.  {\emph{Bottom Left:}}  CH$_3$OH spectrum after despiking and baseline-subtracting.  
   {\emph{Bottom Right:}}  We investigate the significance of both the positive and negative CH$_3$OH response of the full spectrum.  On the y-axis, we define the average of the CH$_3$OH positive and negative response $[\sum\mathrm{T_A^\ast(line)} - \sum\mathrm{T_A^\ast}(\mathrm{inverse})]/2$ (i.e. [the sum of antenna temperature T$_\mathrm{A}^\ast$ over 7 contiguous velocity channels minus the same sum over velocity channels that are 8.63~km~s$^{-1}$ lower] / 2).  The red point denotes the peak of the CH$_3$OH line at $+8$~km~s$^{-1}$.  Black points denote the same calculation for other velocity bins, excluding channels where the calculation would involve part of the bandwidth with spikes, or (for clarity) the line or its negative response.  Hence, these black points show the statistics for a potentially false line due to noise.  The dotted blue lines show the mean $\pm3\sigma$ bounds derived from all of the plotted points.  We find the red point denoting the average CH$_3$OH line and negative response to be the only point that is significant (at the 5$\sigma$ level).}
   \label{fig:nobaseline}%
   \end{sidewaysfigure}

We present an overview of the methods for reducing the CH$_3$OH spectrum and determining the robustness of the CH$_3$OH detection.  Noticeable spikes towards the ends of the band were blanked before the baseline was fit and subtracted from the spectrum (Figure~\ref{fig:nobaseline}).  Both the positive and negative response of the CH$_3$OH line can be seen at a velocity separation of 8.6~km~s$^{-1}$ due to frequency-switching (see Section~\ref{results} for more details).  The robustness of the detection was investigated by averaging the CH$_3$OH positive and negative response and investigating how likely these features are caused by noise (Figure~\ref{fig:nobaseline}).

\section{Full \textsc{RADEX} parameter space}
\label{radex_explained}
\textsc{RADEX} is a one-dimensional non-LTE radiative transfer code that assumes isothermal and homogenous medium without large-scale velocity fields using the escape probability method (i.e. the probability a photon will break out of the surrounding medium).  The program is iterative, finding a solution for the level populations and calculating the radiation temperature $\mathrm{T_R}$ using the following method:
  
\begin{enumerate}

\item Input parameters are: molecular data file from LAMDA (\citealt{2005A&A...432..369S}, which include term energies, statistical weights, Einstein coefficients and rate coefficients for collisional de-excitation), frequency range of the transition, kinetic temperature of the region, number of collision partners (typically H2 as the only collision partner, which has been scaled so that H$_2$O is the collisional partner for our models), H$_2$O density, temperature of the background radiation field, column density of the molecule being modelled and FWHM line width.
\item An initial estimate of the level populations is made by assuming optically thin emission and statistical equilibrium considering the background radiation field (typically 2.73 K blackbody representing the cosmic microwave background or CMB).
\item The optical depths are then calculated for the molecular line.
\item The program iteratively continues to calculate new level populations with new optical depth values until both converge on a consistent solution.
\item The program outputs are: background-subtracted molecular line intensities, excitation temperature and optical depth.
\end{enumerate}

The full parameter space of the \textsc{RADEX} models that fit the CH$_3$OH observation can be found in Table~\ref{full_table_parameters}.

\begin{table}
\centering
\caption{\textsc{RADEX} model parameter space for varying cloud lengths.  `Upper' denotes the upper limits on the column and volume densities.  The smallest cloud length that fits the CH$_3$OH detection is 1750~km, though we include the 1000~km cloud for reference.   $^a$T$_\mathrm{rad}$ is the expected peak main-beam temperature of our CH$_3$OH detection corrected for beam dilution.  $^b$ n$_\mathrm{RADEX}$(H$_2$O) is the range of H$_2$O densities that are fit the CH$_3$OH data.  $^c$T$_\mathrm{exc}$ is the CH$_3$OH excitation temperature range for the models.  $^d$$\tau\mathrm{(CH_3OH)}$ is the optical depth range for CH$_3$OH models. }
\begin{tabular}{c c c c c c c c}
\hline
\hline
\multicolumn{8}{c}{\textsc{RADEX} parameters}\\
\hline
Cloud $l$ & N$_\mathrm{upper}$($\mathrm{H_2O}$) & n$_\mathrm{upper}$($\mathrm{H_2O}$) & n$_\mathrm{upper}$($\mathrm{H_2}$) & T$_\mathrm{rad}$ & n$_\mathrm{RADEX}$(H$_2$O) & T$_\mathrm{exc}$ & $\tau\mathrm{(CH_3OH)}$ \\
(km) & (cm$^{-2}$) & (cm$^{-3}$) & (cm$^{-3}$) & (K) & (cm$^{-3}$) & (K)\\
\hline

(1000) &  6.5$\times10^{18}$ & 6.5$\times10^{10}$ & 6.5$\times10^{11}$ & 351.780 & -- & -- & -- \\
(1250) & 4.1$\times10^{18}$ & 3.3$\times10^{10}$ & 3.3$\times10^{11}$ & 225.163 & -- & -- & -- \\ 
(1500) & 2.9$\times10^{18}$ & 1.9$\times10^{10}$ & 1.9$\times10^{11}$ & 156.383 & -- & -- & -- \\
1750 & 2.1$\times10^{18}$ & 1.2$\times10^{10}$ & 1.2$\times10^{11}$ & 114.911 & 1.5$\times10^9$ -- 1.1$\times10^{10}$ & 100 & 2--13 \\
2000 & 1.6$\times10^{18}$ & 8.1$\times10^{9}$ & 8.1$\times10^{10}$ & 87.995 & 7.9$\times10^8$-- 7.9$\times10^9$ & 100 & 1--12 \\
2250 & 1.3$\times10^{18}$ & 5.7$\times10^{9}$ & 5.7$\times10^{10}$ & 69.540 & 4.5$\times10^8$--5.6$\times10^9$ & 100 & 0.737--9 \\
2500 & 1.0$\times10^{18}$ & 4.1$\times10^{9}$ & 4.1$\times10^{10}$ & 56.340 & 3.2$\times10^8$--4.0$\times10^9$ & 100 & 0.525--7 \\
5000 & 2.6$\times10^{17}$ & 5.2$\times10^{8}$ & 5.2$\times10^{9}$ & 14.135 & 3.2$\times10^7$--5.0$\times10^8$ & 99--100 & 0.106--0.335 \\
7500 & 1.2$\times10^{17}$ & 1.5$\times10^{8}$ &1.5$\times10^{9}$ &6.319 & 7.9$\times10^6$--1.4$\times10^8$ & 97--100 & 0.047--0.141 \\
10000 & 6.5$\times10^{16}$ & 6.5$\times10^{7}$ & 6.5$\times10^{8}$ & 3.583 & 3.2$\times10^6$--6.3$\times10^7$ & 93--100 & 0.026--0.080 \\
12500 & 4.1$\times10^{16}$ & 3.3$\times10^{7}$ & 3.3$\times10^{8}$ & 2.317 & 1.4$\times10^6$--3.2$\times10^7$ & 86--99 & 0.017--0.052 \\
15000 & 2.9$\times10^{16}$ & 1.9$\times10^{7}$ & 1.9$\times10^{8}$ & 1.629 & 7.9$\times10^5$--1.8$\times10^7$ & 78--99 & 0.012--0.038 \\
17500 & 2.1$\times10^{16}$ & 1.2$\times10^{7}$ & 1.2$\times10^{8}$ & 1.214 & 5.6$\times10^5$--1.1$\times10^7$ & 72--98 & 0.009--0.029 \\
20000 & 1.6$\times10^{16}$ & 8.1$\times10^{6}$ & 8.1$\times10^{7}$ & 0.945 & 4.0$\times10^5$--7.9$\times10^6$ & 66--97 & 0.007--0.025 \\
22500 & 1.3$\times10^{16}$ & 5.7$\times10^{6}$ & 5.7$\times10^{7}$ & 0.761 & 2.8$\times10^5$--5.6$\times10^6$ & 60--96 & 0.006--0.021 \\
25000 & 1.0$\times10^{16}$ & 4.1$\times10^{6}$ & 4.1$\times10^{7}$ & 0.629 & 2.5$\times10^5$--4.0$\times10^6$ & 57--94 & 0.005--0.018 \\
27500 & 8.6$\times10^{15}$ & 3.1$\times10^{6}$ & 3.1$\times10^{7}$ & 0.531 & 2.0$\times10^5$--2.8$\times10^6$ & 53--92 & 0.004--0.016 \\
30000 & 7.2$\times10^{15}$ & 2.4$\times10^{6}$ & 2.4$\times10^{7}$ & 0.457 & 1.6$\times10^5$--2.2$\times10^6$ & 49--90 & 0.004--0.015 \\
32500 & 6.1$\times10^{15}$ & 1.9$\times10^{6}$ & 1.9$\times10^{7}$ & 0.399 & 1.4$\times10^5$--1.8$\times10^6$ & 47--88 & 0.003--0.013 \\
35000 & 5.3$\times10^{15}$ & 1.5$\times10^{6}$ & 1.5$\times10^{7}$ &0.353 & 1.3$\times10^5$--1.4$\times10^6$ & 45--85 & 0.003--0.013 \\
37500 & 4.6$\times10^{15}$ & 1.2$\times10^{6}$ & 1.2$\times10^{7}$ & 0.316 & 1.1$\times10^5$--1.1$\times10^6$ & 43--82 & 0.003--0.012 \\
40000 & 4.1$\times10^{15}$ & 1.0$\times10^{6}$ & 1.0$\times10^{7}$ & 0.286 & 1.0$\times10^5$--1.0$\times10^6$ & 41--80 & 0.003--0.012 \\
42500 & 3.6$\times10^{15}$ & 8.4$\times10^{5}$ & 8.4$\times10^{6}$ & 0.261 & 8.9$\times10^4$--7.9$\times10^5$ & 40--77 & 0.003--0.011 \\
45000 & 3.2$\times10^{15}$ & 7.1$\times10^{5}$ &7.1$\times10^{6}$ & 0.240 & 7.9$\times10^4$--7.1$\times10^5$ & 38--74 & 0.003--0.011 \\
47500 & 2.9$\times10^{15}$ & 6.0$\times10^{5}$ & 6.0$\times10^{6}$ & 0.222 & 7.9$\times10^4$--5.6$\times10^5$ & 38--70 & 0.002--0.009 \\
50000 & 2.6$\times10^{15}$ & 5.2$\times10^{5}$ & 5.2$\times10^{6}$ & 0.207 & 7.9$\times10^4$--5.0$\times10^5$ & 38--68 & 0.002-0.009 \\
52500 & 2.4$\times10^{15}$ & 4.5$\times10^{5}$ & 4.5$\times10^{6}$ & 0.194 & 7.1$\times10^4$--4.5$\times10^5$ & 36--66 & 0.002-0.009 \\
55000 & 2.1$\times10^{15}$ & 3.9$\times10^{5}$ & 3.9$\times10^{6}$ & 0.182 & 6.3$\times10^4$--3.5$\times10^5$ & 35--62 & 0.002-0.009 \\
57500 & 2.0$\times10^{15}$ & 3.4$\times10^{5}$ & 3.5$\times10^{6}$ & 0.172 & 5.6$\times10^4$--3.2$\times10^5$ & 34--60 & 0.002-0.009 \\
60000 & 1.8$\times10^{15}$ & 3.0$\times10^{5}$ & 3.0$\times10^{6}$ & 0.164 & 5.6$\times10^4$--2.8$\times10^5$ & 34--58 & 0.002-0.008 \\
62500 & 1.7$\times10^{15}$ & 2.7$\times10^{5}$ & 2.7$\times10^{6}$ & 0.156 & 5.6$\times10^4$--2.5$\times10^5$ & 34--56 & 0.002--0.008 \\
65000 & 1.5$\times10^{15}$ & 2.4$\times10^{5}$ & 2.4$\times10^{6}$ & 0.149 & 5.0$\times10^4$--2.2$\times10^5$ & 32--54 & 0.002-0.008 \\
67500 & 1.4$\times10^{15}$ & 2.1$\times10^{5}$ & 2.1$\times10^{6}$ &0.143 & 5.0$\times10^4$--2.0$\times10^5$ & 32--52 & 0.002-0.008 \\
70000 & 1.3$\times10^{15}$ & 1.9$\times10^{5}$ & 1.9$\times10^{6}$ & 0.138 & 5.0$\times10^4$--1.8$\times10^5$ & 32--50 & 0.002-0.008 \\
72500 & 1.2$\times10^{15}$ & 1.7$\times10^{5}$ & 1.7$\times10^{6}$ & 0.133 & 4.5$\times10^4$ --1.6$\times10^5$ & 31--48 & 0.002--0.008 \\
\hline
\end{tabular}
\label{full_table_parameters}
\end{table}

\end{document}